\documentclass{PoS}
\usepackage{slashed}
\usepackage{graphicx}
\usepackage{amsmath}
\newcommand{\bs}[1]{{\bf #1}}
\title{Mass anomalous dimension of SU$(2)$ with $N_f=8$ using the spectral density method}

\ShortTitle{Mass anomalous dimension of SU$(2)$ with $N_f=8$ using the spectral density method}

\author{\speaker{Joni M. Suorsa}\\
        Department of Physics and Helsinki Institute of Physics, University of Helsinki,
P.O. Box 64, FI-00014 Helsinki, Finland\\
        E-mail: \email{joni.suorsa@helsinki.fi}}

\author{Viljami Leino\\
        Department of Physics and Helsinki Institute of Physics, University of Helsinki}
        
\author{Jarno Rantaharju\\
        $CP^3$-Origins, IFK \& IMADA, University of Southern Denmark and RIKEN Advanced Institute of Computational Science}
        
\author{Teemu Rantalaiho\\
        Department of Physics and Helsinki Institute of Physics, University of Helsinki}
        
\author{Kari Rummukainen\\
        Department of Physics and Helsinki Institute of Physics, University of Helsinki}
        
\author{Kimmo Tuominen\\
        Department of Physics and Helsinki Institute of Physics, University of Helsinki}
 
\author{David J. Weir\\
        University of Stavanger}

\abstract{SU$(2)$ with $N_f=8$ is believed to have an infrared conformal fixed point. We use the spectral density method to evaluate the coupling constant dependence of the mass anomalous dimension for massless HEX smeared, clover improved Wilson fermions with  Schrödinger functional boundary conditions.}

\FullConference{The 33rd International Symposium on Lattice Field Theory\\
		14 -18 July 2015\\
		Kobe International Conference Center, Kobe, Japan*}

\begin{document}

\section{Introduction}
Non-Abelian infrared-conformal gauge theories have been considered as models for physics beyond the Standard Model. In these models the anomalous dimension $\gamma_m$ of the fermion operator $\overline{\psi}\psi$ plays an important role. 
The scaling of the spectral density of the massless Dirac operator is
governed by the mass anomalous dimension \cite{deldebbio}. 
The explicit
calculation of the eigenvalue distribution is costly, but recent advances
in applications of stochastic methods \cite{luscher} have made the mode number
\cite{patella} of the Dirac operator a viable quantity to determine the mass
anomalous dimension from.
The theory which we are studying is SU$(2)$ with $N_f = 8$ fermions in the fundamental representation, which lies within the conformal window \cite{viljami}.

The mode number of the Dirac operator, 
\begin{equation}\label{moden1}
\nu(\Lambda) = 2\int_0^{\sqrt{\Lambda^2 - m^2}} \rho(\lambda)d\lambda,
\end{equation} 
where $\rho(\lambda)$ is the eigenvalue density of the Dirac operator, is known to follow a scaling behaviour\footnote{In lattice units.} 
 of
\begin{equation}\label{modenumber1}
\nu(\Lambda) \simeq \nu_0(m) +  C\left[\Lambda^2 - m^2\right]^{2/(1+\gamma_*)}
\end{equation}
in some intermediate energy range between the infrared and the ultraviolet in the vicinity of the fixed point. Here $\gamma_*$ is the mass anomalous dimension $\gamma_m$ at the fixed point, $\nu_0(m)$ is an additive constant, $C$ is a dimensionless constant, and $m$ is the quark mass. The range where this power law 
behavior holds is not 
known \textit{a priori}, and needs to be determined by trial and error.


The spectral density scaling method has been applied to various models before \cite{degrand,patella2,anna1,anna3,anna4,forcrand,deldebbio2,anna2,cichy,landa,keegan2,keegan}, but in the case of SU$(2)$ the coupling constant dependence of the mass anomalous dimension has not been investigated before.

The mass anomalous dimension can also be obtained by using the Schr\"odinger functional mass step scaling function \cite{stepscaling}. In what follows we will compare results obtained using this method to results obtained using the spectral density method.

\begin{figure}
\begin{center}
\includegraphics[width=0.69\textwidth]{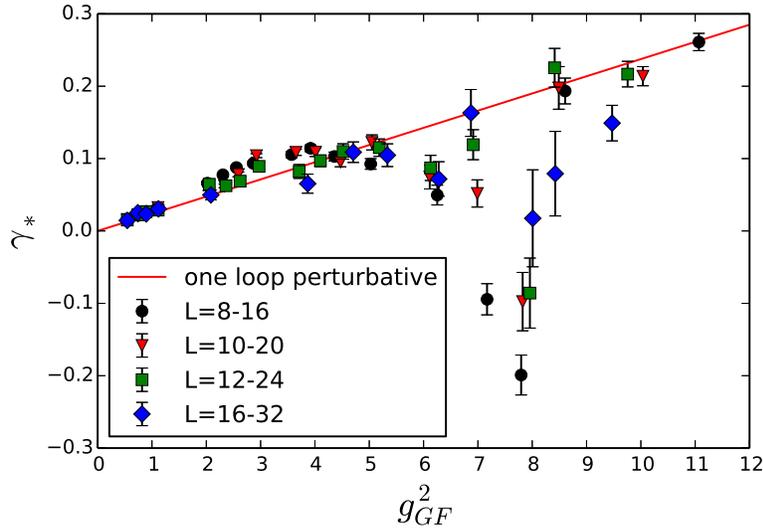}
\label{viljamiplotti}
\end{center}
\caption{The mass anomalous dimension as a function of the gradient flow coupling constant obtained using the mass step scaling function. The different symbols correspond to different lattice size pairings. The results deviate from the perturbative
one loop result at large coupling. The fixed point is at $g_{GF}^2\sim 6$, and the
results for larger couplings are not understood.}
\end{figure}

\section{Mass step scaling}

We simulate SU$(2)$ with $N_f=8$ using HEX smeared \cite{hex}, clover improved \cite{clover} Wilson fermions,
using the same parameters as for the evaluation of the running coupling in \cite{viljami}.
We tune the hopping parameter to $\kappa =\kappa_c$ in order to have zero PCAC quark mass. We simulate the theory at seven different values of $\beta$ corresponding to measured gauge couplings from $g_{GF}^2 = 0.55$ to $g_{GF}^2 = 9.49$ on a $V=32^4$ lattice, where $g_{GF}^2$ is evaluated using the gradient flow step scaling method \cite{viljami,sf}.

For the evaluation of the mass anomalous dimension using the step scaling method, we set the spatial gauge links to unity at temporal boundaries:
\begin{equation}
 U_i(\mathbf{x},t=0) = U_i(\mathbf{x},t=L) = \mathbf{1}.
\end{equation}
The mass anomalous dimension $\gamma_m$ is measured
from the running of the pseudoscalar density renormalization constant
\cite{stepscaling,dellamorte}
\begin{align}
Z_P(L) = \frac{\sqrt{3 f_1} }{f_P(L/2)},
\label{Zp}
\end{align}
where
\begin{align}
 f_P(t) &= \frac{-a^6}{3L^6}\sum_{\bs y,\bs z}
 \langle P^a(\bs x, t)
 \,\bar\zeta(\bs y)\gamma_5\frac12\sigma^a\zeta(\bs z)\rangle, \\
 f_1 &= \frac{-a^{12}}{3 L^{12}} \sum_{\bs u,\bs v, \bs y, \bs z}
 \langle\bar \zeta'(\bs u)\gamma_5\frac12 \sigma^a \zeta'(\bs v)\,
  \bar\zeta(\bs y)\gamma_5\frac12 \sigma^a\zeta(\bs z)\rangle.
\end{align}
Here $P^a(x) = \overline{\psi}(x)\gamma_5 \frac{1}{2}\sigma^a \psi(x)$, and $\zeta$ and $\zeta'$ are boundary quark sources at $t=0$ and $t=L$ respectively.
Now we can define the mass step scaling function as \cite{stepscaling}
\begin{align}
\Sigma_P(u,s,L/a) &=
   \left. \frac {Z_P(g_0,sL/a)}{Z_P(g_0,L/a)} \right |_{g_{GF}^2(g_0,L/a)=u}
   \label{Sigmap}\\
\sigma_P(u,s) &= \lim_{a/L\rightarrow 0} \Sigma_P(u,s,L/a).
\end{align}
We choose $s=2$ and find the continuum step scaling function $\sigma_P$ by
measuring $\Sigma_P$ at $L/a=8$, $10$, $12$ and $16$.
The mass anomalous dimension can then be obtained from the mass step scaling function \cite{dellamorte}.
Denoting the function estimating the anomalous dimension $\gamma_m(u)$ by
$\gamma_*(u)$, we have
\begin{align}
  \gamma_*(u) = -\frac{\log \sigma_P(u,s)}{\log s }.
\label{eq:gammastar}
\end{align}

Our preliminary results are shown in Fig. \ref{viljamiplotti}. The method gives results comparable to one loop perturbation theory predictions at small gauge coupling $g_{GF}^2$, but deviates from the perturbative results at large coupling as
the theory flows toward the fixed point at $g_{GF}^2 \sim 6$ \cite{viljami}. 


\section{Spectral density method}
We calculate the mode number per unit volume of Eq. \ref{moden1} by using 
\begin{equation}
\nu(\Lambda) =\lim_{V\to \infty}  \frac{1}{V}\left< \textnormal{tr }\mathbb{P}(\Lambda)\right>,
\end{equation}
where the operator $\mathbb{P}(\Lambda)$ projects from the full eigenspace of $M = m^2 - \slashed{D}^2$ to the eigenspace of eigenvalues lower than $\Lambda^2$, and the trace is calculated stochastically. 

We use the lattices obtained from the step scaling analysis, and use between 12 to 20 well separated configurations for each value of the gauge coupling. We calculate the mode number for 100 values of $\Lambda^2$ ranging from $10^{-4}$ to $0.3$.


We expected the two constants $\nu_0(m)$ and $m^2$ in Eq. \ref{modenumber1} to be negligible since we have tuned the quark mass $m_{PCAC}$ to zero and the additive constant $\nu_0(m)$ is related to the part of the spectrum that feels the effects of the nonzero mass. In principle the unknown renormalisation factor in $m=Z_A m_{PCAC}$ forbids simply setting these two constants to zero. In practice we observed the two constants to be negligible: in our analysis we used
\begin{equation}\label{modenumber2}
\nu(\Lambda) \simeq  C\Lambda^{4/(1+\gamma_*)}
\end{equation}
and checked that the error relative to the form including all four parameters, Eq. \ref{modenumber1}, was $\mathcal{O}(10^{-3})$.

The fit range was determined by varying the lower and the upper limit of the fit range and observing the stability of the fit and the parameter values and their errors. As a cross reference we compared the value of $\gamma_*$ obtained using the spectral density method for small couplings to the value obtained using the step scaling method in order to further assess wether the chosen fit range was good or not.

\begin{figure}
\begin{center}
\includegraphics[width=0.69\textwidth]{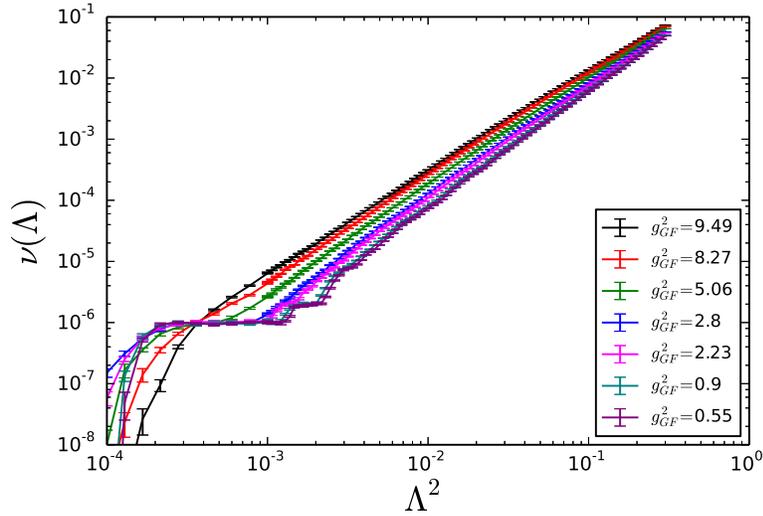}
\caption{The mode number calculated for different gauge couplings on a $V = 32^4$ lattice.}
\label{all_beta}
\end{center}
\end{figure}

In Fig. \ref{all_beta} we present the mode number data we have calculated. It is apparent that the smaller coupling simulations suffer from finite size effects which manifest in the step-like structure of the mode number curve as $\Lambda \to 0$, but this disappears at couplings $g_{GF}^2 \ge 2.8$.

\begin{figure}
\begin{center}
\includegraphics[width=0.69\textwidth]{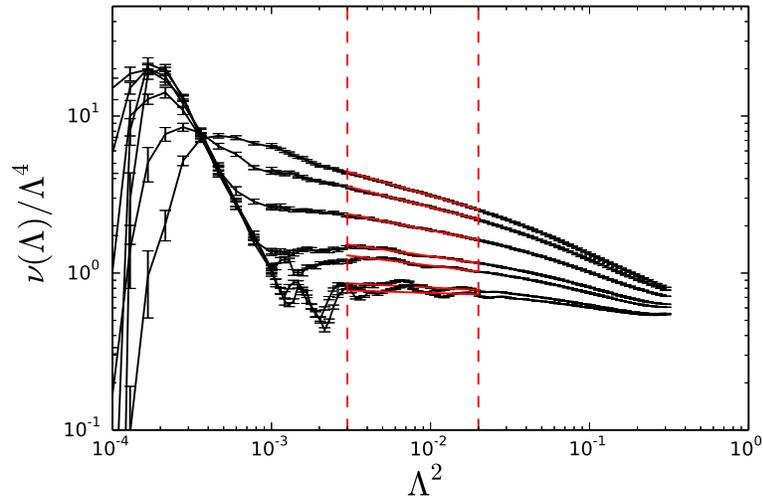}
\caption{The mode number divided by $\Lambda^4$ as a function of $\Lambda^2$. The fit function and the fit range are indicated by solid and dashed red lines respectively. The curves are in the order of descending gauge coupling.}
\label{fit_range}
\end{center}
\end{figure}

In Fig. \ref{fit_range} we plot the mode number divided by the fourth power of the eigenvalue scale as a function of the eigenvalue scale squared with the chosen fit range and the fit function of Eq. \ref{modenumber2} shown overlaid in red. The curves are in the order of descending gauge coupling. It is clear that the chosen fit range for smaller coupling values, which appear as the lowest curves in the plot, goes into the step-like structure of the mode number curve, but it is a relatively good approximation of the average behaviour of the curve. When the data is presented in this way, we expect in our massless case that there is a region where the curves are linear, which is the correct window for the fit.

\begin{figure}
\begin{center}
\includegraphics[width=0.69\textwidth]{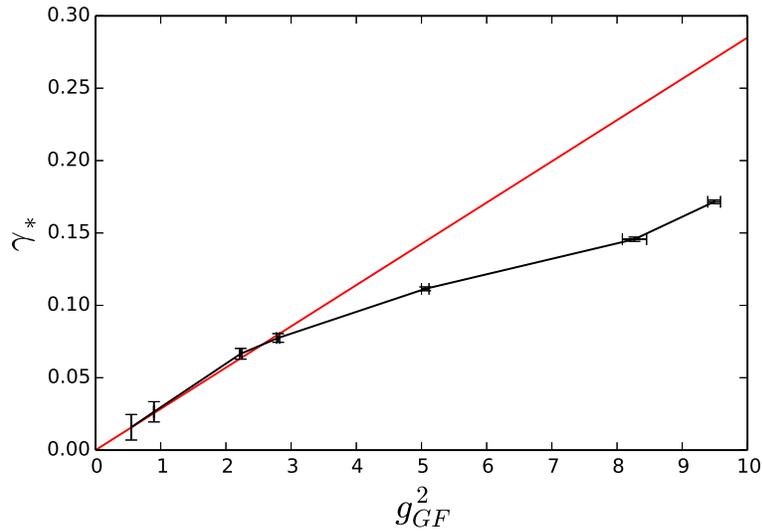}
\caption{The value of $\gamma_*$ obtained by fitting Eq. \protect\ref{modenumber2} to the data in Fig. \protect\ref{all_beta} is shown with black points and the one loop perturbative result with a red line.}
\label{result}
\end{center}
\end{figure}

Our main result is shown in Fig. \ref{result} where we plot the mass anomalous dimension $\gamma_*$ obtained from fitting Eq. \ref{modenumber2} to the data as a function of the gauge coupling $g_{GF}^2$. 
In a similar fashion to the results obtained using 
the mass step scaling method shown in Fig. \ref{viljamiplotti}, the spectral density 
method seems 
to give results comparable with the one loop perturbative prediction for small gauge 
coupling values. But whereas the mass step scaling method showed highly nontrivial behaviour in our simulations at gauge coupling 
values above $g_{GF}^2 \sim 6$, the spectral density method gives 
results that exhibit consistent behaviour with increasing coupling.

\section{Conclusions}


We have determined the mass anomalous dimension of SU(2) gauge theory with eight Dirac fermions in the fundamental representation of the gauge group using the spectral density method. We have demonstrated that the method gives results compatible with perturbation theory and with nonperturbative mass step scaling method at weak coupling.
As the coupling increases, the results were observed to deviate from the perturbative result.  

A major source of error that is not easily quantifiable is the choice of the fit range where Eq. \ref{modenumber2} (or Eq. \ref{modenumber1}) is used to describe the data. Our choice of the fit range was guided by our preliminary results using the mass step scaling method and requirements on the stability and quality of the fit. The results at small coupling suffer from sensitivity to the variation of the fit range, but this is not a problem at couplings where perturbation theory can be trusted as matching can be made reliably. However, when comparing with the results from the mass step scaling method, the fitting procedure would benefit from smaller lattice artefacts and well established setting of the fit ranges.
The larger coupling results were largely insensitive to the choice of the fit range, and consequently it seems that the observed behaviour at large coupling is a genuine nonperturbative feature, and not an artefact due to fit uncertainties.

\section*{Aknowledgements}
J.M.S. is supported by the Jenny and Antti Wihuri foundation. K.R., V.L., and K.T. are supported by the Academy of Finland
grants 267842, 134018, and 267286. J.R. is supported by the Danish
National Research Foundation grant number DNRF:90 and by a Lundbeck Foundation Fellowship grant. T.R. is supported by the Magnus Ehrnrooth foundation, and D.J.W. is supported by the People Programme (Marie Sk{\l}odowska-Curie actions) of the European Union Seventh Framework Programme (FP7/2007-2013) under grant agreement number PIEF-GA-2013-629425. The simulations were performed at the Finnish IT Center for Science (CSC) in Espoo, Finland, on the Fermi supercomputer at Cineca in Bologna, Italy, and on the K computer at Riken AICS in Kobe, Japan. Parts of the simulation program have been derived from the MILC lattice simulation program \cite{milc}.


\begin{thebibliography}{99}


\bibitem{deldebbio}
L. Del Debbio and R. Zwicky,
\newblock Phys.Rev. D82 (2010) 014502 (arXiv:1005.2371 [hep-ph])


\bibitem{luscher} 
L. Giusti and M. L\"uscher,
\newblock JHEP 0903 (2009) 013 (arXiv:0812.3638 [hep-lat])

\bibitem{patella}
A. Patella,
\newblock Phys.Rev. D84 (2011) 125033 (arXiv:1106.3494 [hep-th])


\bibitem{viljami}
V. Leino, T. Karavirta, J. Rantaharju, T. Rantalaiho, K. Rummukainen, J.M. Suorsa, and K. Tuominen,
\newblock These proceedings


\bibitem{degrand}
T. DeGrand,
\newblock Phys.Rev. D80 (2009) 114507 (arXiv:0910.3072 [hep-lat])

\bibitem{patella2}
A. Patella,
\newblock Phys.Rev. D86 (2012) 025006 (arXiv:1204.4432 [hep-lat])

\bibitem{anna1}
A. Hasenfratz, A. Cheng, G. Petropoulos, and D. Schaich,
\newblock PoS LATTICE2012 (2012) 034 (arXiv:1207.7162 [hep-lat])

\bibitem{anna3}
A. Cheng, A. Hasenfratz, and D. Schaich,
\newblock Phys.Rev. D85 (2012) 094509 (arXiv:1111.2317 [hep-lat])


\bibitem{anna4}
A. Cheng, A. Hasenfratz, G. Petropoulos, and D. Schaich,
\newblock JHEP 1307 (2013) 061 (arXiv:1301.1355 [hep-lat])

\bibitem{forcrand}
P. de Forcrand, S. Kim, and W. Unger,
\newblock JHEP 1302 (2013) 051 (arXiv:1208.2148 [hep-lat])


\bibitem{deldebbio2}
L. Del Debbio, B. Lucini, C. Pica, A. Patella, A. Rago, and S. Roman,
\newblock PoS LATTICE2013 (2014) 067 (arXiv:1311.5597 [hep-lat])

\bibitem{anna2}
A. Cheng, A. Hasenfratz, G. Petropoulos, and D. Schaich,
\newblock PoS LATTICE2013 (2014) 088 (arXiv:1311.1287 [hep-lat])

\bibitem{cichy}
K. Cichy,
\newblock JHEP 1408 (2014) 127 (arXiv:1311.3572 [hep-lat])

\bibitem{landa}
D. Landa-Marban, W. Bietenholz, and I. Hip,
\newblock Int.J.Mod.Phys. 25 (2014) 1450051 (arXiv:1307.0231 [hep-lat])



\bibitem{keegan2}
M. García Pérez, A. González-Arroyo, L. Keegan, and M. Okawa,
\newblock JHEP 1508 (2015) 034 (arXiv:1506.06536 [hep-lat])

\bibitem{keegan}
L. Keegan,
\newblock These proceedings (arXiv:1508.01685 [hep-lat])



\bibitem{stepscaling}
S. Capitani,  M. L\"uscher, R. Sommer, and H. Wittig,
\newblock Nucl.Phys. B544 (1999) 669-698 (arXiv:hep-lat/9810063v3)



\bibitem{hex}
S. Capitani, S. Durr, and C. Hoelbling, 
\newblock JHEP 0611 (2006) 028 (arXiv:hep-lat/0607006)


\bibitem{clover}
K. Jansen and C. Liu, 
\newblock Comput.Phys.Commun. 99 (1997) 221

\bibitem{sf}
M. L\"uscher, R. Narayanan, P. Weisz, and U. Wolff,
\newblock Nucl.Phys. B384 (1992) 168-228 (arXiv:hep-lat/9207009)


\bibitem{dellamorte}
M. Della Morte, R. Hoffmann, F. Knechtli, J. Rolf, R. Sommer, I. Wetzorke, and U. Wolff,
\newblock Nucl. Phys. B729 (2005) 117-134 (arXiv:hep-lat/0507035)

\bibitem{milc}
http://physics.utah.edu/\textasciitilde detar/milc.html


\end{thebibliography}
\end{document}